# Theoretical study of electronic and atomic structures of $(MnO)_n$

Hiori Kino,[a] Lucas K. Wagner[b] and Lubos Mitas[c]

[a]National Institute for Materials Science, 1-2-1 Sengen, Tsukuba, Ibaraki 305-0047, Japan.
[b]Berkeley Nanoscience and Nanoengineering Institute, University of California at Berkeley, Berkeley, CA 94720.
[c]Department of Physics, North Carolina State University, Raleigh, NC 27695-8202.

We calculate the electronic and atomic structure of $(MnO)_n$ (n=1-4) using the HF exchange, VWN, PBE and B3LYP exchange-correlation functionals. We also perform diffusion Monte Carlo calculation to evaluate more accurate energies. We compare these results and discuss the accuracy of the exchange-correlation functionals.

## 1. Introduction

Nanoparticles have attracted attention because their exotic properties are not only scientifically interesting but also technologically important. Among them, manganese oxide clusters are particularly attractive, since they exhibit ferromagnetic behavior in experiments,[1,2] in contrast to the antiferromagnetic one in bulk manganese oxide. This kind of nano-sized ferromagnetic particles may be applied to high-density storage devices, which utilize the magnetic behavior of materials. In this context, it is interesting to study the geometry of small manganese oxide clusters in connection with their magnetic properties using first-principle methods.[3-5]

Similarly in biophysics, nano-sized manganese oxide clusters play a number of important roles. For example, the photosystem II in green plants contains a small calcium manganese oxide cluster in its oxygen evolving center.[6] Theoretical studies of this system have been carried out[7] using the B3LYP functional without questioning the accuracy of the method.

The results of density functional theory (DFT) studies depend on the choice of an exchange-correlation (XC) functional, and it is usually possible to establish which XC functional is the most appropriate for a given system of interest. On the other hand, it is widely known that electronic and magnetic structures calculated by DFT methods show a number of serious shortcomings and errors when it comes to strongly correlated systems. It is, therefore, important to employ alternative approaches in order to verify whether a given XC functional is sufficiently accurate to describe such systems correctly.

In this paper, we study the atomic and magnetic structure of $(MnO)_n$ for $n \leq 4$. We compare the results obtained using several XC functionals. In addition, we employ the diffusion quantum Monte Carlo (DMC) method, which is capable of capturing both correlation and exchange effects with high accuracy, subject only to the fixed-node approximation.[8] Since the fixed-node bias is of the order of merely 5-10 percent of the correlation energy, the energy differences such as binding and excitation energies usually agree with the experimental ones within a few percent. In this study, we use a triple zeta basis set with diffuse functions. We also employ the effective core potential for Mn (Ne core) and O (He core). We use gaussian03 package for the unrestricted Hartree-Fock (HF), SVWN (Slater exchange and VWN[9,10] correlation functional also called LSDA), PBE[11] and B3LYP[12] calculations, whereas QWalk package[13] is used for our DMC calculations. The trial function is a single Slater determinant multiplied by the two-body part of the Jastrow function reported in Ref. 14. In most cases, we evaluate DMC total energies using the B3LYP geometries. In some cases, we also perform the DMC calculation with the equilibrium geometries of other XC functionals, using the wavefunctions obtained from the B3LYP functional since the B3LYP gives better d-p hybridizations which decrease the nodal errors.[14-16]

For simplicity, we consider only ferromagnetic states (in equilibrium geometries, the energy difference between the ferromagnetic state and the corresponding antiferromagnetic state is expected to be of the order of $10^2$ K, and we don't discuss such small differences in this report). The purpose of this study is to clarify the atomic and magnetic structures of $(MnO)_n$ ($n \leq 4$) and verify the accuracy of different XC functionals through comparisons with the DMC results.

## 2. Calculations and Results
### 2.1 General Trends

We calculate the binding energies of $(MnO)_n$ per n (defined as $BE_n = -\frac{1}{n}(E((MnO)_n) - n\,E(Mn) - n\,E(O))$, where $E(X)$ is the total energy of X) as a function of spin multiplicity ($M$) with each XC functional. The general trend is as follows. As expected, SVWN overestimates while HF underestimates the value of $BE_n$ in all cases. The PBE functional results in the slightly smaller value, whereas the B3LYP functional leads to the slightly larger value of $BE_n$ when compared to the DMC results. In comparison with B3LYP and DMC, SVWN and PBE tend to predict smaller values of $M$ in the ground state. Considering the DMC total energies of geometries optimized using different XC functionals, the B3LYP geometries lead to the lowest total energies, and hence the largest $BE_n$, while the PBE geometries give the second largest $BE_n$ in many cases.

### 2.2 MnO

Wagner and Mitas have already studied this system using the same method as here,[14-16] but with a different form of the Jastrow function: we repeat these calculations for the purpose of consistency and completeness. We summarize the calculated results in Table 1. The experimental equilibrium length ($r_e$) of MnO is 1.65 Å,[17,18] which the B3LYP functional reproduces very well (1.64 Å). PBE leads to almost the same value (1.63Å), but SVWN underestimates it (1.60Å). The DMC energy evaluated with the B3LYP geometry is within the statistical errors of the previous report in the DMC optimization ( $\cong$ 1.65 Å) when we take into account of the small difference in    .

Table 1(a) shows $r_e$ as a function of spin multiplicity ($M$). We find that $r_e$ depends on $M$. $r_e$ increases as $M$ increases (except in the HF calculation). This dependence indicates that the correct value of $M$ is crucial in order to obtain an accurate geometry.

Fig. 1(b) shows the binding energies. With all XC functionals, the value of $M=6$ corresponds to the ground state geometry, except for the HF case. Table 1(b) shows that binding energies strongly depend on the choice of the XC functional. The $BE_1$s, 1.35, 6.12, 4.90, 3.55 and 3.67±0.02eV, are obtained from the HF, SVWN, PBE, B3LYP and DMC calculations respectively.

In the ground state, the binding energy calculated using the B3LYP functional is compatible with the DMC result and the experimental value (3.85±0.08eV[18]). The energy difference is larger for other spin multiplicities (excited states). For example, the B3LYP binding energy for $M=4$ is about one half of the DMC value, and the difference in binding energy between the B3LYP and DMC results is larger in $(MnO)_n$ for $n \geq 2$ as shown later.

## 2.3 $(MnO)_2$

Table 2 shows $BE_2$s, and Fig. 1 also displays $BE_2$ and some of the geometries. The largest $BE_2$s, 1.88, 7.82, 6.21, 5.14 and 5.61±0.02eV, are obtained from the HF, SVWN, PBE, B3LYP and DMC calculations respectively. The spin multiplicity $M=9$ is computed using SVWN and PBE in the most stable structure as reported by Nayak and Jena.[3,4] In contrast, the value of 11 corresponds to the ground state geometry in the HF, B3LYP and DMC calculations.

Now we set $M=11$ and study bond lengths and angles. The B3LYP functional leads to a geometry characterized by 1.90 Å ($r$(MnO)) and $95^0$ ($\angle$(OMnO)). The PBE functional results in similar values (1.89 Å and $95^0$, respectively). The SVWN functional underestimates slightly (1.86 Å and $94^0$, respectively). As we mentioned above, the value of $M$ in the ground state computed with the B3LYP functional is different from that obtained using the PBE and SVWN functionals. With the latter functionals, the value of $M=9$ leads to the ground state geometries described by 1.81Å and $97^0$ respectively in the PBE case, whereas the ground state geometry has slightly smaller bond length (1.78 Å and $98^0$) in the SVWN case.

## 2.4 $(MnO)_3$

Table 3 shows the $BE_3$s, while Fig. 2 displays $BE_3$s and some of the stable structures. The most stable geometry is circular.[3,4] The largest $BE_3$s, 3.05, 8.68, 7.19, 6.12 and 6.72±0.03eV, are obtained from the HF, SVWN, PBE, B3LYP and DMC calculations respectively. With the SVWN and PBE functional, the value of $M=14$ corresponds to the ground state geometry, while the value of 16 is obtained in the HF, B3LYP and DMC calculations. When $M=16$, the B3LYP functional leads to the geometry characterized by the bond length and angles of 1.86 Å ($r$(MnO)), $108^0$ ($\angle$(MnOMn)) and $132^0$ ($\angle$(OMnO)). With the SVWN and PBE functionals, the optimal geometries when $M=12$, 14 and 16 have almost the same energy.

## 2.5 $(MnO)_4$

There are three stable high-symmetry geometry types: circular, cubic and ladder-like (we abbreviate them as circle, cube and ladder in Fig. 3 and Table 4, where the equilibrium geometry type for a given M is shown in parenthesis). Table 4 shows the $BE_4$s, while Fig. 3 depicts $BE_4$s and some of the geometries. The most stable structure is circular in all cases. The largest $BE_4$s, 9.17, 7.48, 6.45 and 7.04±0.03eV, are obtained from the SVWN, PBE, B3LYP and DMC calculations respectively. The value of $M$=19 leads to the most stable geometry with the SVWN and PBE functionals, whereas the value of $M$=21 results in the most stable structure in the B3LYP and DMC calculations. The second most stable structure is cubic with the SVWN and B3LYP functionals, but it is ladder-like with the PBE functional. When $M$=21, the B3LYP calculation results in the geometry featuring 1.848 Å ($r$(MnO)), 120.5$^0$ ($\angle$(MnOMn)) and 149.5$^0$ ($\angle$(OMnO)).

Next let us consider structures when $M$=23. The most stable geometry with the B3LYP functional is distorted and drastically different from the high-symmetry circular one obtained from the PBE calculation as shown in Fig. 3(b). The SVWN functional also leads to a similar high-symmetry circular geometry. In the DMC calculation, the high-symmetry circular geometry results in much smaller $BE_4$ than the distorted geometry obtained from the B3LYP functional.

When $M$=21, we find that the ladder structure is unstable in the B3LYP calculation. On the other hand, the SVWN and PBE calculations lead to stable ladder geometries, but these have smaller $BE_4$s than the most stable circular geometry obtained using the B3LYP functional (evaluated in the DMC calculation). Finally, we note that, during optimizations, we find a very low-symmetry structure with $M$=15 and the SVWN functional (Fig. 3(b)), the $BE_4$ of which is 9.09eV. This is slightly larger than the $BE_4$ at $M$=17(circle) (9.06). (The values are taken from the SVWN calculation.)

## 3. Discussion

Now let us compare theoretical results with experimental ones. Experiments have clarified the geometry of MnO and $(MnO)_2$.[17-20] We start with the case of MnO. It is known that the DMC calculation results in the bond length that agrees with the experimental one.[14-16] We showed that the B3LYP and PBE functionals best reproduce the experimental equilibrium length. The situation is different with $(MnO)_2$. The experimental bond lengths and angle are 2.0Å ($r$(MnO)), 2.60Å ($r$(MnMn)), and 100$^0$.($\angle$(OMnO)).[20] B3LYP underestimates the bond lengths by 0.05~0.1Å. The SVWN and PBE functionals generate much smaller and less realistic bond lengths. The larger discrepancy in the SVWN and PBE cases may be connected to the smaller $M$: In general, SVWN and PBE functionals lead to smaller multiplicities than B3LYP and DMC in the ground state. The SVWN and PBE multiplicities are 6, 9, 14 and 19, while B3LYP and DMC lead to the values of 6, 11, 16 and 21 for n=1, 2, 3 and 4 respectively. We can't compare computed geometries for $n$>2 or the values of $M$s with experimental ones because to the best of our knowledge, there is no experiment for $(MnO)_n$ (n>2). The result of geometry comparison for $(MnO)_n$ n=1 and 2 suggests that we must employ SVWN and PBE carefully to $(MnO)_n$ clusters, which are known to be strongly correlated systems.

We evaluated the DMC binding energies in the geometries obtained using different XC functionals. Let us determine which functional is the most accurate by comparing the binding energies. As discussed above, the B3LYP functional leads to the same $M$ as DMC, and the B3LYP geometry has the lowest DMC energy in the ground state. In the exicited state ($M$=23), B3LYP also performs well with the $(MnO)_4$ cluster. Here the B3LYP functional leads to the distorted circular geometry, while the SVWN and PBE result in the high-symmetry one. When one carries out the DMC calculation, the B3LYP geometry is found to lead to the lowest energy. We conclude that the B3LYP functional generates the most accurate energy surfaces.

In DFT structure optimization, there are possibly lower energy geometries than the ones we have found. For example, we identified a low-symmetry structure of $(MnO)_4$ when $M$=15, that has lower energy than the cubic and ladder structures when evaluated using the SVWN functional. Already at $n$=4, it is difficult to find a global minimum since the optimization becomes difficult and complex. For larger systems, one needs to employ more sophisticated techniques that are able to explore the configuration space efficiently and can find the global minimum in an automatic fashion.[21]

The geometries we studied are obtained with the XC functionals. We can't exclude the possibility that structure optimization in other DFT calculations as well as in the DMC one could lead to other geometries that have lower energies than those considered in this study. Nayak and Jena[3,4] have reported that the ladder structure is the most stable $(MnO)_4$ geometry in calculation using the BPW91 functional[9]. However, our calculation shows that the most stable structure is circular, regardless of the choice of XC functional. The difference in the ground state geometry comes from the choice of the XC functional. This result illustrates that the properties of manganese oxide clusters are sensitive to the XC functional, emphasizing the importance of accurate treatment of exchange-correlation terms.

We have shown that the B3LYP functional generates the best energy surfaces, but there exist small discrepancies. The B3LYP calculation underestimates $BE_n$ with respect to the DMC one by 3% for MnO and 9% for $(MnO)_n$ (n=2-4). The differences in $BE_n$ of the excited states are larger, and there is also a difference in the bond lengths of the $(MnO)_2$ cluster as discussed above. Yet it may be possible to find XC functional suitable to these clusters.

## 4. Conclusions

In summary, we have studied the atomic and electronic structures of $(MnO)_n$ clusters for n=1-4 obtained from the HF, SVWN, PBE, B3LYP and DMC calculations. In MnO, the PBE, B3LYP and DMC calculations accurately reproduce the experimental geometry. Nevertheless, all functionals underestimate the bond length in $(MnO)_2$ than the experimental one, although the B3LYP functional result in the closest. We have found a new theoretical ground state geometry for $(MnO)_4$. We compared the results with the DMC calculation in order to establish the relative accuracy of the XC functionals. The calculated spin multiplicities in the ground states are the same in DMC and B3LYP calculations, while the SVWN and PBE result in slightly smaller values of spin multiplicies for $(MnO)_n$ ($n \geq 2$). We conclude that the B3LYP functional is the most accurate method for treating $(MnO)_n$ clusters.

## Acknowledgments
This research has been partially supported (L.M.) by the DOE DE-FG05-08OR23336 and NSF EAR-0530110 grants, by INCITE and CNMS initiative at ORNL, and by a Grant-in-Aid for Scientific

Research in Priority Areas "Development of New Quantum Simulators and Quantum Design" (No.17064017) of The Ministry of Education, Culture, Sports, Science, and Technology, Japan. We thank Dr. M. Todorovic for useful discussions.

**Table 1.** (a) Equilibrium lengths ($r_e$) in Å and (b) Binding energies in eV as a function of spin multiplicity ($M$) and the exchange-correlation functional (HF, SVWN, PBE, B3LYP) in MnO. We evaluate the DMC binding energies using the geometry obtained with the B3LYP functional. The Monte-Carlo errors in the DMC calculation are in the parentheses.

(a)

| $M$ | HF | SVWN | PBE | B3LYP |
|---|---|---|---|---|
| 2 | 1.64 | 1.52 | 1.55 | 1.57 |
| 4 | 1.90 | 1.57 | 1.61 | 1.61 |
| 6 | 1.86 | 1.60 | 1.63 | 1.64 |
| 8 | 1.98 | 1.92 | 1.96 | 1.82 |
| 10 | 3.27 | | | |

(b)

| $M$ | HF | SVWN | PBE | B3LYP | DMC |
|---|---|---|---|---|---|
| 2 | -6.96 | 3.98 | 2.59 | 1.23 | |
| 4 | -3.41 | 4.93 | 3.63 | 1.30 | 2.18(2) |
| 6 | 1.01 | 6.12 | 4.90 | 3.55 | 3.67(2) |
| 8 | 1.35 | 3.44 | 2.89 | 2.26 | 2.65(3) |
| 10 | -1.10 | | | | |

**Table 2.** The DMC binding energies (in eV per MnO unit) as a function of $M$ calculated for geometries obtained using different exchange-correlation functionals (SVWN, PBE, B3LYP) in $(MnO)_2$. We employ B3LYP orbitals as the trial wavefunctions in the DMC calculation. The cell in gray shows the largest value.

| $M$ | SVWN | PBE | B3LYP |
|---|---|---|---|
| 9 | 4.73(2) | 4.87(2) | 5.02(3) |
| 11 | 5.58(3) | 5.58(2) | 5.61(2) |
| 13 | | | 4.39(2) |

**Table 3.** The DMC binding energies (in eV per MnO unit) as a function of $M$ calculated for geometries obtained using different exchange-correlation functionals (SVWN, PBE, B3LYP) in $(MnO)_3$. We employ B3LYP orbitals as the trial wavefunctions in the DMC calculation. The cell in gray shows the largest value.

| $M$ | SVWN | PBE | B3LYP |
|---|---|---|---|
| 14 | 5.99(3) | 5.80(3) | 6.12(2) |
| 16 | 6.41(3) | 6.63(2) | 6.72(3) |
| 18 | | | 5.58(1) |

**Table 4.** The DMC binding energies (in eV per MnO unit) as a function of $M$ calculated for geometries obtained using different exchange-correlation functionals (SVWN, PBE, B3LYP) in $(MnO)_4$. We employ B3LYP orbitals as the trial wavefunctions in the DMC calculation. We note that the $M$=21(ladder) is unstable with the B3LYP functional. The cell in gray shows the largest value.

| M(geometry) | SVWN | PBE | B3LYP |
|---|---|---|---|
| 19(circle) | 6.45(3) | 6.54(3) | 6.53(2) |
| 21(circle) | 6.98(3) | 7.01(2) | 7.04(3) |
| 23(circle) | 5.61(3) | 5.63(3) | 6.31(2) |
| 19(cube) | | | 6.23(2) |
| 21(cube) | | | 6.78(2) |
| 23(cube) | | | 6.05(2) |
| 21(ladder) | 6.48(3) | 6.73(2) | |

**Figure 1**.
(a)

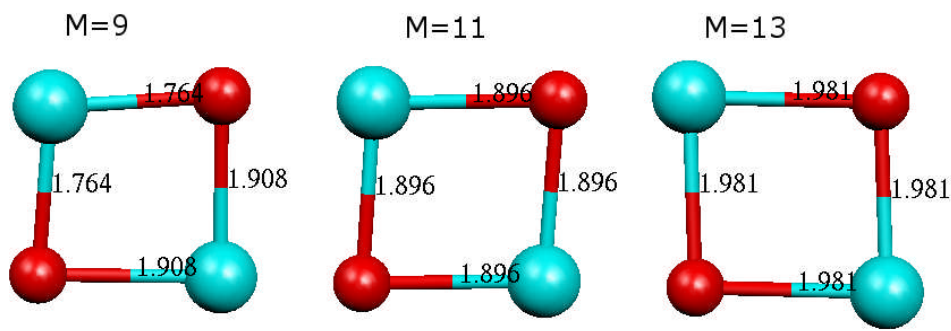

(b)

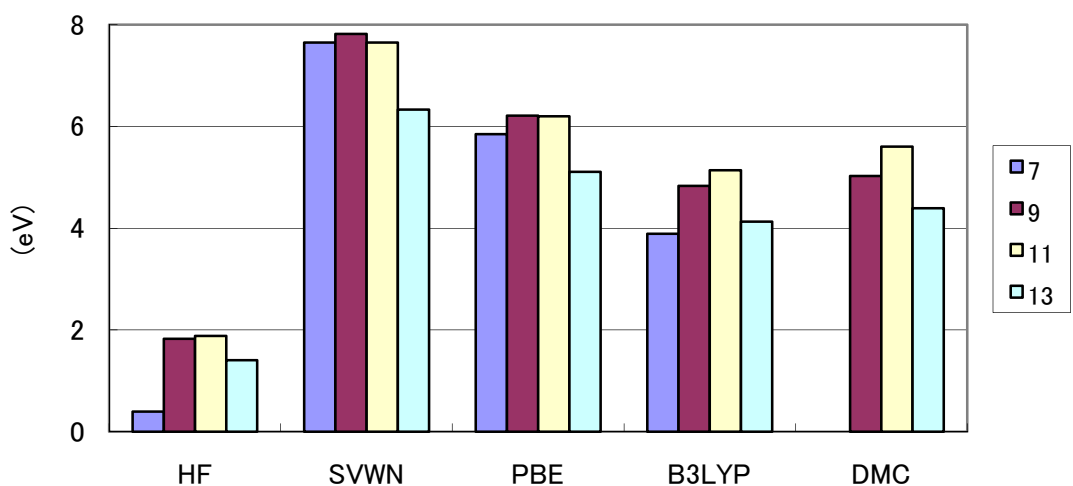

(a) Equilibrium geometries of (MnO)$_2$ obtained from the B3LYP exchange-correlation functional when spin multiplicities ($M$) =9, 11 and 13. Large purple spheres denote manganese and while the small red ones correspond to oxygen. (b) Binding energies (in eV per MnO unit) as a function of $M$ using different exchange-correlation functionals (HF, SVWN, PBE, B3LYP) for (MnO)$_2$. We evaluate the DMC binding energies using the geometries obtained from the B3LYP functional.

Figure 2.

(a)

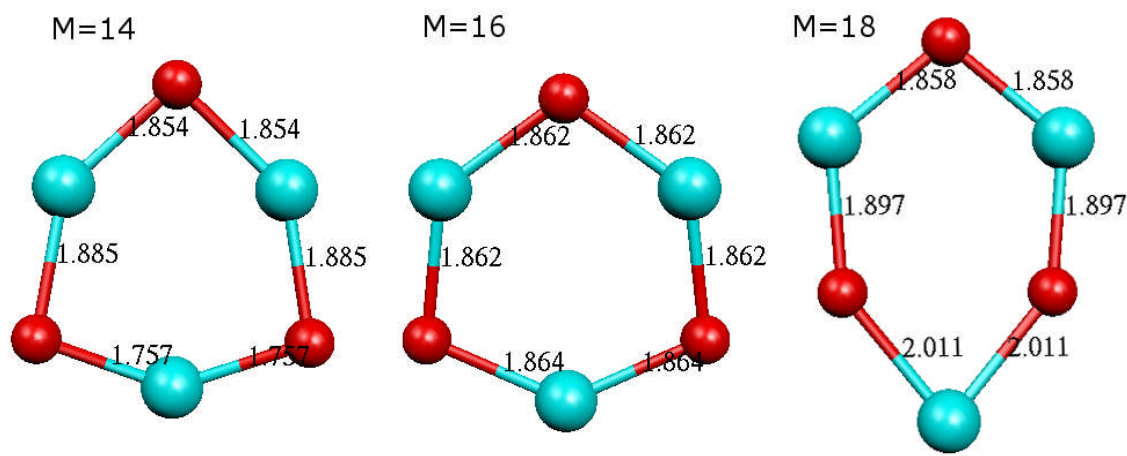

(b)

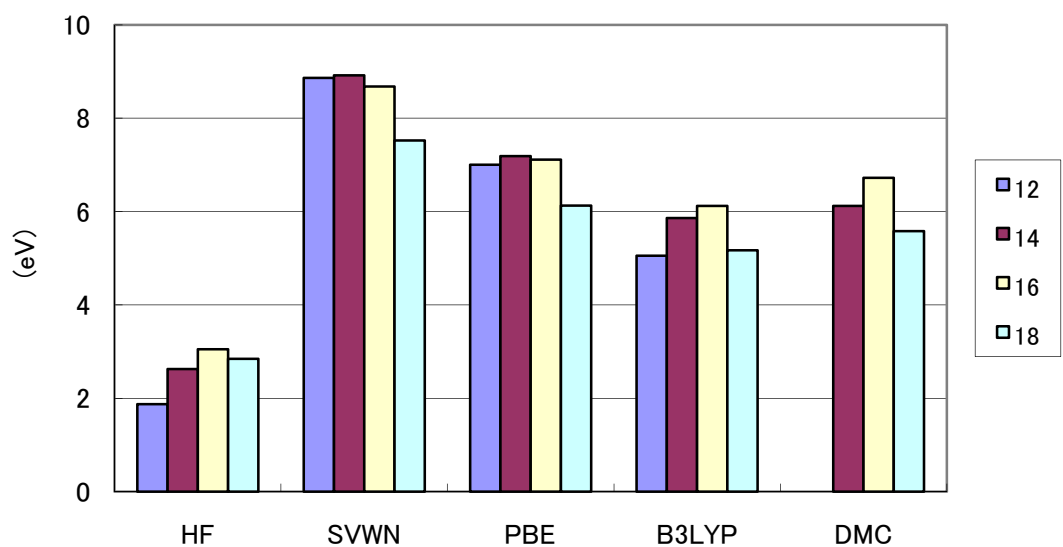

(a) Equilibrium geometry of $(MnO)_3$ obtained from the B3LYP exchange-correlation functional at $M$=14, 16 and 18. (b) Binding energies (eV per MnO unit) as a function of $M$ using different exchange-correlation functionals (HF, SVWN, PBE, B3LYP) in $(MnO)_3$. We evaluate the DMC binding energies using the geometries obtained from the B3LYP functional.

Figure 3.

(a)

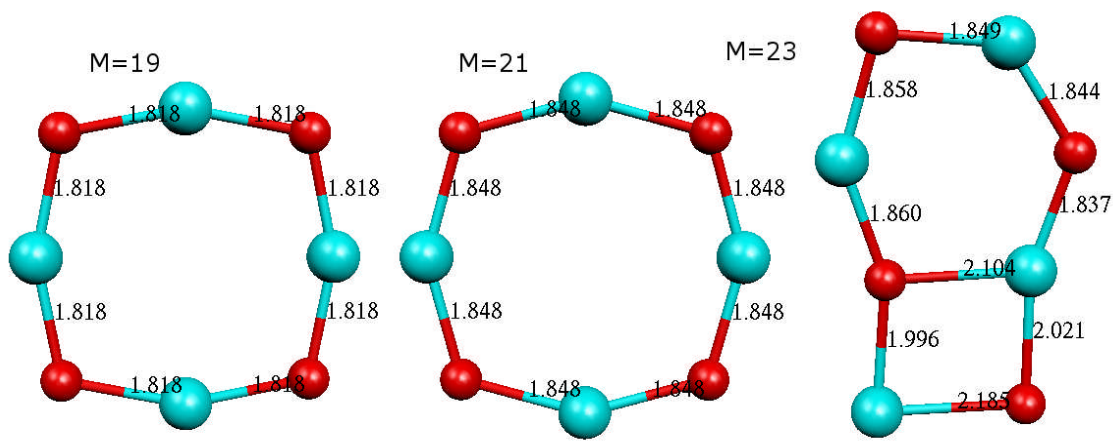

(b)

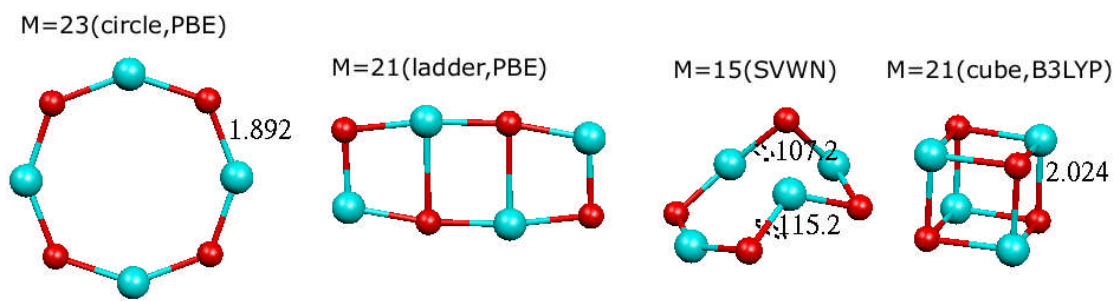

(c)

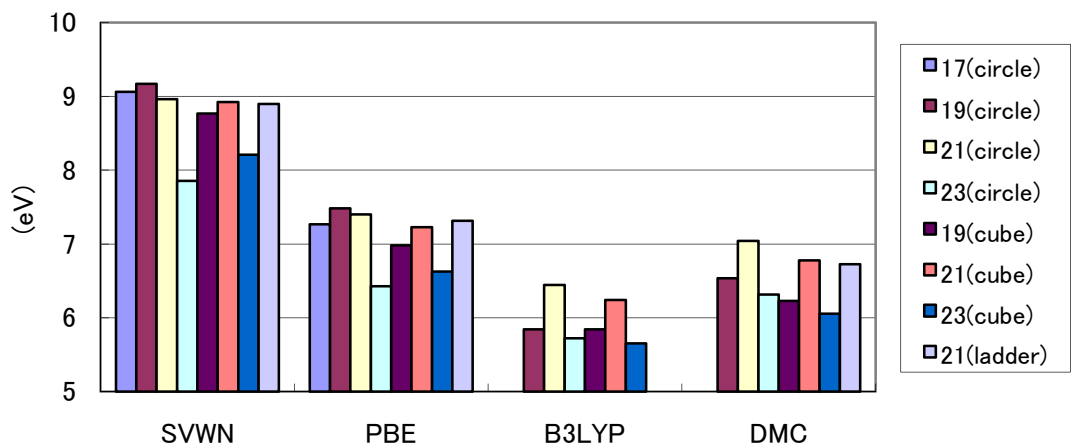

(a) Equilibrium geometry of (MnO)$_4$ obtained from the B3LYP exchange-correlation functional. (b) Equilibrium geometry of (MnO)$_4$ with the exchange-correlation functionals and at some $M$s. (c) Binding energies (eV per MnO unit) as a function of $M$ using different exchange-correlation functionals (SVWN, PBE, B3LYP). We evaluate the DMC binding energies using the geometries obtained from the B3LYP functional, but the DMC energy when $M$=21(ladder) is obtained using the PBE geometry, because the geometry is unstable with the B3LYP functional.